\documentclass[referee]{aa} 
\usepackage{graphicx}
\def\lesssim{\mathrel{\hbox{\rlap{\hbox{\lower4pt\hbox{$\sim$}}}\hbox{$<$}}}}
\def\gtrsim{\mathrel{\hbox{\rlap{\hbox{\lower4pt\hbox{$\sim$}}}\hbox{$>$}}}}
\newcommand{\mincir}{\raise -2.truept\hbox{\rlap{\hbox{$\sim$}}\raise5.truept
\hbox{$<$}\ }}
\newcommand{\magcir}{\raise -2.truept\hbox{\rlap{\hbox{$\sim$}}\raise5.truept
\hbox{$>$}\ }}
\newcommand{\siml}{\raise -2.truept\hbox{\rlap{\hbox{$\sim$}}\raise5.truept
\hbox{$<$}\ }}
\newcommand{\simg}{\raise -2.truept\hbox{\rlap{\hbox{$\sim$}}\raise5.truept
\hbox{$>$}\ }}
\newcommand{\be}{\begin{equation}}
\newcommand{\ee}{\end{equation}}
\newcommand{\ba}{\begin{eqnarray}}
\newcommand{\ea}{\end{eqnarray}}
\newcommand {\h} {$\mathrm{h^{-1}}$ Mpc $\;$}

\newcommand {\hh} {$\mathrm{h^{-1}}$ Mpc}
\newcommand {\ks} {km~s$^{-1} \;$}
\newcommand {\kss} {km~s$^{-1}$}
\newcommand {\msun} {$\mathrm{h^{-1} \  M_{\odot} \;}$}

\begin{document}
   \title{Structure and Evolution of Galaxy Clusters: 
Internal Dynamics of ABCG\,209 at z$\sim$0.21.
\thanks{Based on
 observations collected at the European Southern Observatory, Chile
(Proposal ESO N° 68.A-0116)}}

%
\author{A. Mercurio\inst{1} 
\and M. Girardi\inst{1}
\and W. Boschin\inst{1}
\and P. Merluzzi\inst{2}
\and G. Busarello\inst{2}}
   \offprints{A. Mercurio}

\institute{Dipartimento di Astronomia, Universit\`{a} degli Studi di Trieste,
Via Tiepolo 11, I-34100 Trieste, Italy\\
\email{mercurio,girardi,boschin@ts.astro.it}
\and   INAF - Osservatorio Astronomico di Capodimonte,
via Moiariello 16, I80100- Napoli, Italy\\
\email{merluzzi,busarello@na.astro.it} 
}
   \date{Received 24-07-02 /  accepted 11-10-02}

\abstract{
We study the internal dynamics of the rich galaxy cluster ABGC\,209 on the
basis of new spectroscopic and photometric data.
The distribution in redshift shows that ABCG\,209 is a well isolated
peak of 112 detected member galaxies at $\mathrm{z=0.209}$, characterised
by a high value of the line--of--sight velocity dispersion,
$\mathrm{\sigma_v=1250}$--$1400$ \kss, on the whole observed area (1 \h
from the cluster center), that leads to a virial mass of
$\mathrm{M=1.6}$--$2.2\times 10^{15}$\msun within the virial radius, 
assuming the dynamical equilibrium.
The presence of a velocity gradient in the velocity field, the
elongation in the spatial distribution of the colour--selected likely
cluster members, the elongation of the X--ray contour levels in the
Chandra image, and the elongation of cD galaxy
show that ABCG\,209 is characterised by a preferential NW--SE
direction.
We also find a significant deviation of the velocity distribution from a
Gaussian, and relevant evidence of substructure and
dynamical segregation.
All these facts show that ABCG\,209 is a strongly evolving cluster,
possibly in an advanced phase of merging.

\keywords{Galaxies: clusters: general --
Galaxies: clusters: individual: ABCG\,209 -- Galaxies: distances and 
redshifts -- intergalactic medium -- Cosmology: observations}}
\titlerunning{Internal Dynamics of ABCG\,209 at z$\sim$0.21}
   \maketitle
%

\section{Introduction}

The investigation of clusters of galaxies offers a rare possibility to link
many aspects of astrophysics and cosmology and, in particular, to
understand the processes that lead to the formation of structures.

In hierarchical clustering cosmological scenarios galaxy clusters form
from the accretions of subunits. Numerical simulations show that
clusters form preferentially through anisotropic accretion of
subclusters along filaments (West et al. \cite{wes91}; Katz \& White
\cite{kat93}; Cen \& Ostriker \cite{cen94}; Colberg et al \cite{col98},
\cite{col99}). The signature of this anisotropic cluster formation is
the cluster elongation along the main accretion filaments
(e.g., Roettiger et al. \cite{roe97}).
Therefore the knowledge of the properties of galaxy clusters,
plays an important role in the study of large--scale
structure (LSS) formation and in constraining cosmological models. 
 
By studying the structure of galaxy clusters it is possible to discriminate
between different cosmological models (e.g., Richstone et
al. \cite{ric92}; Lacey \& Cole \cite{lac94}; Thomas et
al. \cite{tho98}).  In fact, in a low--density universe the clustering
tends to freeze at (z+1) = $\mathrm{\Omega_{M}^{-1}}$, while in a
high--density universe it continues to grow to the present day. This
implies that clusters in a low--density universe are expected to be
dynamically more relaxed and to have less subsystems, called
substructures.

On small scales, clusters appear as complex systems involving a
variety of interacting components (galaxies, X--ray emitting gas, dark
matter). A large fraction of clusters (30\%-40\%) contain substructures, as
shown by optical and X--ray studies (e.g.,  Baier \& Ziener
\cite{bai77}; Geller \& Beers \cite{gel82}; Girardi et
al. \cite{gir97}; Jones \& Forman \cite{jon99}) and by recent results
coming from the gravitational lensing effect (e.g.,  Athreya et
al. \cite{ath02}; Dahle et al. \cite{dah02}), suggesting that they are still in the
dynamical relaxation phase. Indeed, there is growing evidence that
these subsystems arise from merging of groups and/or clusters
(cf. Buote \cite{buo02}; and Girardi \& Biviano \cite{gir02} for reviews).

Very recently, it was also suggested that the presence of radio halos and
relics in clusters is indicative of a cluster merger. Merger shocks, with 
velocities larger than 10$^3$ km s$^{-1}$, convert a fraction of the
shock energy into acceleration of pre--existing relativistic particles
and provide the large amount of energy necessary for magnetic
field amplification (Feretti \cite{fer00}). This mechanism has been
proposed to explain the radio halos and relics in clusters (Brunetti
et al. \cite{bru01}).

The properties of the brightest cluster members (BCMs) are related to
the cluster merger.  Most BCMs are located very close to the center of
the parent cluster. In many cases the major axis of the BCM is
aligned along the major axis of the cluster and of the surrounding LSS
(e.g., Binggeli \cite{bin82}; Dantas et al. \cite{dan97}; Durret et
al. \cite{dur98}). These properties can be explained if BCMs form by
coalescence of the central brightest galaxies of the merging
subclusters (Johnstone et al. \cite{joh91}).

The optical spectroscopy of member galaxies is the most powerful tool
to investigate the dynamics of cluster mergers, since it provides
direct information on the cluster velocity field.  However this is
often an ardue investigation due to the limited number of galaxies
usually available to trace the internal cluster velocity.  To date, at
medium and high redshifts (z $\gtrsim$ 0.2), only few clusters are really
well sampled in the velocity space (with $>$ 100 members; e.g., 
Carlberg et al \cite{car96}; Czoske et al. \cite{czo02}).

In order to gain insight into the physics of the cluster formation, we
carried out a spectroscopic and photometric study of the cluster
ABCG\,209, at z$\sim$0.2 (Kristian et al. \cite{kri78}; Wilkinson \&
Oke \cite{wil78}; Fetisova \cite{fet81}), which is a rich, very X--ray
luminous and hot cluster (richness class $\mathrm{R=3}$, Abell et
al. \cite{abe89}; $\mathrm{L_X(0.1-2.4\ 
keV) \sim 14\;h_{50}^{-2}\;10^{44}}$ erg $\mathrm{s^{-1}}$,
Ebeling et al. \cite{ebe96}; $\mathrm{T_X\sim10}$ keV, Rizza et
al. \cite{riz98} ) . The first evidence for its complex dynamical
status came from the significant irregularity in the X--ray emission
with a trimodal peak (Rizza et al. \cite{riz98}).  Moreover,
Giovannini et al. (\cite{gio99}) have recently found evidence for the
presence of a possible extended Radio--emission.

The paper is organised as follows. In Sect.~2 we present the new
spectroscopic data and the data reduction. The derivation of the
redshifts is presented in Sect.~3. In Sect.~4 we analyse the dynamics of the
cluster, and in Sect.~5 we complete the dynamical analysis with the
information coming either from optical imaging or from X--ray data.  
In Sect.~6 we discuss the results in terms of two pictures of the
dynamical status of ABCG\,209. Finally, a summary of the main results is
given in Sect.~7.

Throughout the paper, we assume a flat cosmology with $\mathrm{\Omega_M=0.3}$
and $\mathrm{\Omega_{\Lambda}=0.7}$.  For the sake of simplicity in rescaling, we
adopt a Hubble constant of 100 h \ks Mpc$^{-1}$. In this assumption, 1
arcmin corresponds to $\sim$ 0.144 Mpc.  Unless otherwise stated, we
give errors at the 68\% confidence level (hereafter c.l.).

\section{Observations and data reduction}

The data were collected at the ESO New Technology Telescope (NTT) with 
the ESO Multi Mode Instrument (EMMI) in October 2001.

\subsection{Spectroscopy}

Spectroscopic data have been obtained with the multi--object
spectroscopy (MOS) mode of EMMI. Targets were randomly selected
by using preliminary R--band images (T$\mathrm{_{exp}}$=180 s) to construct the
multislit plates. We acquired five masks in four fields (field of view
$5^\prime \times 8.6^\prime$), with different position angles on the
sky, allocating a total of 166 slitlets (alligment stars included).
In order to better sample the denser cluster region, we covered this
region with two masks and with the overlap of the other three masks.
We exposed the masks with integration times from 0.75 to 3 hr with
the EMMI--Grism\#2, yielding a dispersion of $\sim2.8$ \AA/pix and a
resolution of $\sim 8$ \AA FWHM, in the spectral range 385 -- 900 nm.

Each scientific exposure (as well as flat fields and calibration
lamps) was bias subtracted. The individual spectra were extracted and
flat field corrected.  Cosmic rays were rejected in two steps. First,
we removed the cosmic rays lying close to the objects by interpolation
between adjacent pixels, then we combined the different exposures by
using the IRAF\footnote{IRAF is distributed by the National Optical
Astronomy Observatories, which are operated by the Association of
Universities for Research in Astronomy, Inc., under cooperative
agreement with the National Science Foundation.} task IMCOMBINE with
the algorithm CRREJECT (the positions of the objects in different
exposures were checked before).  Wavelength calibration was obtained
using He--Ar lamp spectra. The typical r.m.s. scatter around the
dispersion relation was $\sim$ 15 \ks.  The positions of the objects
in the slits were defined interactively using the IRAF package
APEXTRACT.  The exact object position within the slit was traced in
the dispersion direction and fitted with a low order polynomial to
allow for atmospheric refraction. The spectra were then sky subtracted
and the rows containing the object were averaged to produce the
one--dimensional spectra.

The signal--to--noise ratio per pixel of the one--dimensional galaxy
spectra ranges from about 5 to 20 in the region 380--500 nm.

\subsection{Photometry}

A field of $9.2^\prime \times 8.6^\prime$ (1.2 $\times$ 1.1
$\mathrm{h^{-2}}$ Mpc$^2$ at z=0.209) was observed in B--, V-- and
R--bands pointed to the center of the cluster. Two additional adjacent
fields were observed in V--band in order to sample the cluster at
large distance from the center (out to $\sim$ 1.5 \hh).  The reduction
of the photometric data will be described elsewhere (Mercurio et
al. \cite{mer02}, in preparation).  In order to derive magnitudes and
colours we used the software SExtractor (Bertin \& Arnouts
\cite{ber96}) to measure the Kron magnitude (Kron \cite{kro80}) in an
adaptive aperture equal to $\mathrm{a \cdot r_K}$, where $\mathrm{r_K}$
is the Kron radius and a is a constant. Following Bertin \& Arnouts
(\cite{ber96}), we chose $\mathrm{a=2.5}$, for which it is expected that the
Kron magnitude encloses $\sim 94 \%$ of the total flux of the source.
We use the photometric data to investigate the colour segregation in
Sect.~5.

\section{Redshifts measurements}
 
Redshifts were derived using the cross--correlation technique (Tonry
\& Davis \cite{ton81}), as implemented in the RVSAO package.  We
adopted galaxy spectral templates from Kennicutt (\cite{ken92}),
corresponding to morphological types EL, S0, Sa, Sb, Sc, Ir.  The
correlation was computed in the Fourier domain.

Out of the 166 spectra, 112 turned out to be cluster members (seven of
which observed twice), 22 are stars, 8 are nearby galaxies, 1 is
foreground and 6 are background galaxies. In 10 cases we could not
determine the redshift.

In order to estimate the uncertainties in the redshift measurements,
we considered the error calculated with the cross--correlation
technique, which is based on the width of the peak and on the
amplitude of the antisymmetric noise in the cross correlation.  The
wavelength calibration errors ($\sim$ 15 \ks) turned out to be
negligible in this respect.  The errors derived from the
cross--correlation are however known to be smaller than the true
errors (e.g., Malumuth et al.  \cite{mal92}; Bardelli et
al. \cite{bar94}; Quintana et al. \cite{qui00}). We checked the error
estimates by comparing the redshifts computed for the seven galaxies
for which we had duplicate measurements.  The two data sets agree with
one--to--one relation, but a reasonable value of $\chi^2$ for the fit
was obtained when the errors derived from the cross--correlation were
multiplied by a correction factor ($\sim 1.75$).  A similar correction
was obtained by Malumuth et al.  (\cite{mal92}; 1.6), Bardelli et
al. (\cite{bar94}; 1.87), and Quintana et al. (\cite{qui00}; 1.57).
External errors, which are however not relevant in the study of
internal dynamics, cannot be estimated since only two previous
redshifts are available for ABCG\,209.  

The catalogue of the spectroscopic sample is presented in Table
~\ref{catalogue}, which includes: identification number of each
galaxy, ID (Col.~1); right ascension and declination (J2000),
$\alpha$ and $\delta$ (Col.~2 and ~3); V magnitude (Col.~4); B--R
colour (Col.~5); heliocentric corrected redshift $\mathrm{z}$ (Col.~6) with
the uncertainty $\mathrm{\Delta z}$ (Col.~7);

Our spectroscopic sample is 60\% complete for V$<$20 mag and drops
steeply to 30\% completeness at V$<$21 mag; these levels are reached both
for the external and the internal cluster regions. The spatial
distribution of galaxies with measured redshifts does not show any
obvious global bias. Only around the brightest cluster
member (cD galaxy; c.f. La Barbera et al. \cite{lab02} in preparation)
the spatial coverage is less complete because of geometrical restrictions.
This does not affect our dynamical
analysis because we investigate structures at scales larger than
$\sim$ 0.1 \h.

\section{Dynamical analysis}

\subsection{Member selection and global properties}

ABCG\,209 appears as a well isolated peak in the redshift space. The
analysis of the velocity distribution based on the one--dimensional
adaptive kernel technique (Pisani \cite{pis93}, as implemented by
Fadda et al. \cite{fad96} and Girardi et al. \cite{gir96}) confirms
the existence of a single peak at $\mathrm{z\sim 0.209}$. 

\begin{figure*}
\centering
\includegraphics[width=16cm]{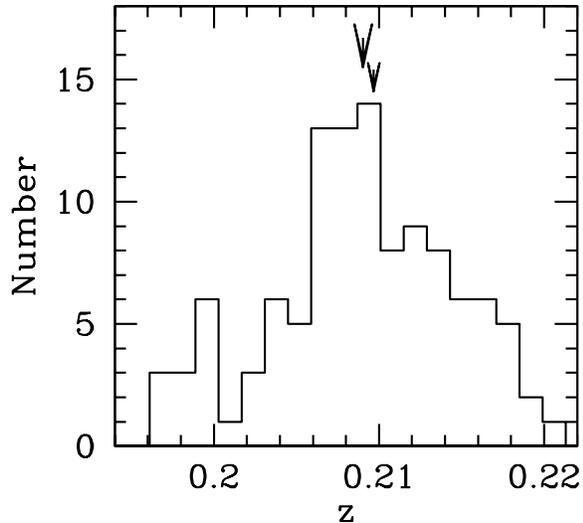}
\vspace{-8cm}
\caption
{Distribution of redshifts of the
cluster members.  The big and small arrows indicate the mean cluster
redshift and the redshift of the cD galaxy, respectively.
}
\label{fighisto}
\end{figure*}

Fig.~\ref{fighisto} shows the redshift distribution of the 112 cluster
members. The mean redshift in the present sample is $\mathrm{<z>=0.2090\pm 0.0004}$,
as derived by the biweight estimator (Beers et al. \cite{bee90}).

In order to determine the cluster center, we applied the
two--dimensional adaptive kernel technique to galaxy positions. The
center of the most dense peak ($\alpha$= 01 31 52.70, $\delta$= -13 36
41.9) is close to the position of cD galaxy ($\alpha$= 01 31 52.54,
$\delta$= -13 36 40.4).

We estimated the line--of--sight (LOS) velocity dispersion, 
$\mathrm{\sigma_v}$, by using the
biweight estimator (ROSTAT package; Beers et al. \cite{bee90}). By applying 
the relativistic correction and the usual correction for velocity errors
(Danese et al. \cite{dan80}), we obtained
$\mathrm{\sigma_v=1394^{+88}_{-99}}$ \kss, where errors were estimated with the
bootstrap method. 

\begin{figure*}
\hspace{1cm}
\vspace{-2cm}
\includegraphics[width=16cm]{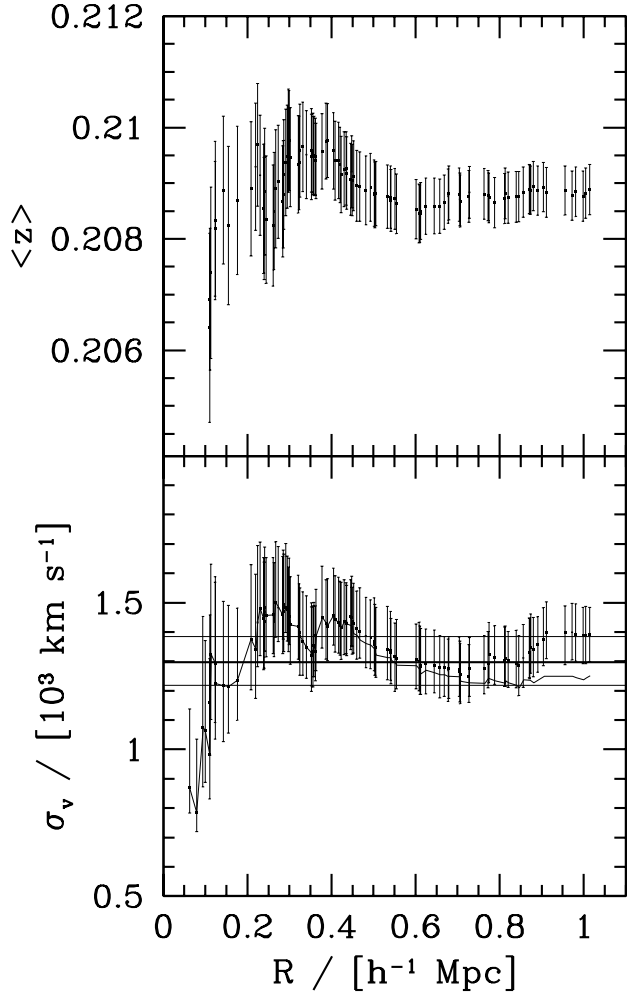}
\caption
{Integrated mean redshift and LOS velocity dispersion profiles (upper
and lower panel, respectively), where the mean and dispersion at a
given (projected) radius from the cluster center is estimated by
considering all galaxies within that radius. The error bars at the
$68\%$ c.l. are shown. In the lower panel, the faint line gives the
profile after the rejection of seven possible
interlopers according to the ``shifting gapper'' method (see text in
Sect.~4.2), and the horizontal lines represent X--ray temperature with
the respective errors (cf. Sect.~5.3) transformed in $\mathrm{\sigma_v}$ imposing
$\mathrm{\beta_{spec}=1}$, where $\mathrm{\beta_{spec}=\sigma_v^2/(kT/\mu m_p)}$, with
$\mathrm{\mu}$ the mean molecular weight and $\mathrm{m_p}$ the proton mass. }
\label{figprof}
\end{figure*}

In order to check for possible variation of $\mathrm{<z>}$ and $\mathrm{\sigma_v}$ with
increasing radius we plot the integrated  mean velocity and 
velocity dispersion profiles in Fig.~\ref{figprof}. The measure of
mean redshift and velocity dispersion sharply vary in the internal
cluster region although the large associated errors do not allow to
claim for a statistically significant variation. On the other hand, 
in the external cluster regions, where the number of galaxies is larger,
the estimates of $\mathrm{<z>}$ and $\mathrm{\sigma_v}$ are quite robust.

Assuming that the system is in dynamical equilibrium, the value of
$\mathrm{\sigma_v}$ leads to a value of the radius of the collapsed,
quasi--virialized region $\mathrm{R_{vir}\sim 1.78}$ \h (cf. Eq.~(1) of Girardi
\& Mezzetti 2001). Therefore our spectroscopic data sample about half
of the virial region ($\sim$ 1 \hh).
Under the same assumption, we estimated the mass of the system by
applying the virial method. In particular, for the surface term
correction to the standard virial mass we assumed a value of $20\%$
(cf. Girardi \& Mezzetti \cite{gir01}), obtaining 
$\mathrm{M(<R_{vir})=2.25^{+0.63}_{-0.65}\times10^{15}}$ \msun.

\subsection{Possible contamination effects}

We further explored the reliability of $\mathrm{\sigma_v}$ related to the
possibility of contamination by interlopers.

\begin{figure*}
\hspace{1cm}
\vspace{-8cm}
\includegraphics[width=16cm]{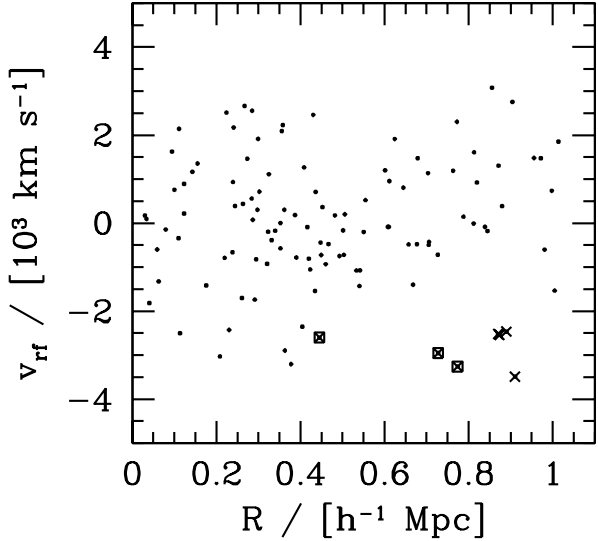}
\caption
{Velocity in the cluster rest frame vs. (projected) clustercentric
distance for the 112 selected members.  The standard application of
the ``shifting gapper'' method would reject galaxies as cluster
members indicated by squares, and a more conservative application also
galaxies indicated by crosses.
}
\label{figvd}
\end{figure*}

The cluster rest--frame velocity vs. projected clustercentric distance
is shown in Fig.~\ref{figvd}.
Although no obvious case of outliers is present, we applied the
procedure of the ``shifting gapper'' by Fadda et al. (\cite{fad96}).
This procedure rejects galaxies that are too far in velocity from the
main body of galaxies of the whole cluster within a fixed bin,
shifting along the distance from the cluster center.  According
to the prescriptions in Fadda et al. (\cite{fad96}), we used a gap of
$1000$ \ks and a bin of 0.4 \hh, or large enough to include at least 15
galaxies.

In this way we rejected three galaxies (indicated by
squares in Fig.~\ref{figvd}). However, for ABCG\,209 the results are
too much sensitive to small changes of the adopted parameters. For
instance, with a bin of $0.5$ \h no galaxy was rejected, while seven
galaxies were rejected with a bin of $0.3$ \h (cf. crosses in
Fig.~\ref{figvd}). In the last case, we obtained a value of
$\mathrm{\sigma_v=1250^{+84}_{-98}}$ \kss, which is slightly smaller than-- (although 
fully consistent with--) the value computed in Sect.~4.1 (cf. also the velocity
dispersion profile in Fig.~\ref{figprof}).
With this value of the velocity dispersion, the computation for the mass
within $\mathrm{R_{vir}=1.59}$ \h leads to a value of total mass
$\mathrm{M(<R_{vir})=1.62^{+0.48}_{-0.46}\times10^{15}}$ \msun.

\begin{figure*}
\hspace{1cm}
\vspace{-8cm}
\includegraphics[width=16cm]{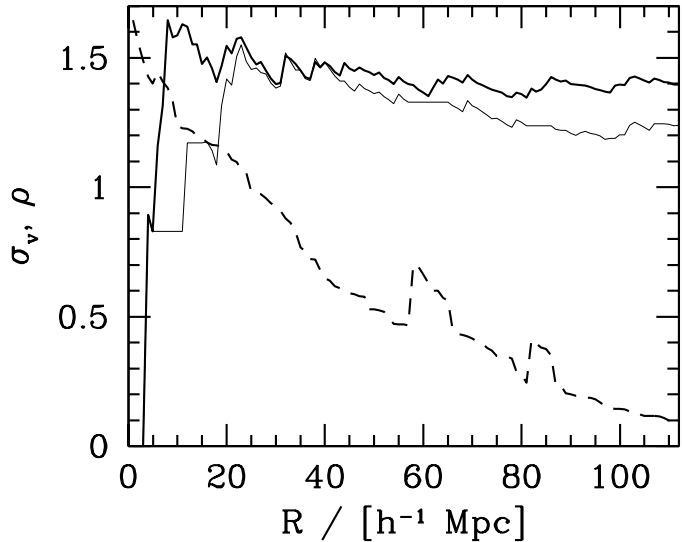}
\caption
{Integrated LOS velocity dispersion profile $\mathrm{\sigma_v}$ (solid line, in
units of $10^3$ \kss) and galaxy number density $\rho$ (dashed line, in
arbitrary units) along the sequences beginning with the center of the
cluster.  The thinner line gives the $\mathrm{\sigma_v}$--profile after the
rejection of
galaxies belonging to the two small density peaks.}
\label{figkittler}
\end{figure*}

We determined the galaxy density and the integrated LOS velocity
dispersion along the sequence of galaxies with decreasing density
beginning with the cluster center
(Kittler \cite{kit76}; Pisani \cite{pis96}). 
As shown in Fig.~\ref{figkittler} the density profile has only two
minor peaks, possibly due to non complete sampling, and their
rejection leads to a small variation in the estimate of $\mathrm{\sigma_v}$ and
in the behaviour of velocity dispersion profile (cf. with Figure~2 of
Girardi et al. \cite{gir96} where the large effect caused by a close
system is shown).

We conclude that the contamination by interlopers cannot explain the
high value of velocity dispersion, which is therefore connected to the
peculiarity of the internal dynamics of the cluster itself. In fact, a
high value of $\mathrm{\sigma_v}$ is already present in the internal cluster
region, namely within $\sim$ 0.2--0.3 \h (cf. Fig.~\ref{figprof}), where the
contamination is expected to be negligible.

\subsection{Velocity distribution}

In order to better characterise the velocity distribution, we
considered its kurtosis $K$, skewness $S$, scaled
tail index $STI$, and the probability associated to the
W--test $P(W)$, (cf. Shapiro \& Wilk \cite{sha65}).

For the kurtosis and the skewness we found $K=2.61\pm0.45$,
$S=-0.20\pm0.23$, respectively, i.e. values that are consistent with a
Gaussian distribution (reference value $K$=3, $S$=0).  The $STI$
indicates the amount of the elongation in a sample relative to the
standard Gaussian. This is an alternative to the classical kurtosis
estimator, based on the data set percentiles as determined from the
order statistics (e.g., Rosenberger \& Gasko \cite{ros83}).  The
Gaussian, which is by definition neutrally elongated, has
$STI=1.0$. Heavier--tailed distributions have $STI \sim$ 1.25 (see,
e.g., Beers et al. \cite{bee91}).  For our data $STI=1.254$,
indicating a heavier--tailed distribution, with 95\%-99\% c.l.
(cf. Table~2 of Bird \& Beers \cite{bir93}).  On the other hand, the
W--test rejected the null hypothesis of a Gaussian parent
distribution, with only a marginal significance at $92.3\%$ c.l.

\begin{figure*}
\hspace{1cm}
\vspace{-8cm}
\includegraphics[width=16cm]{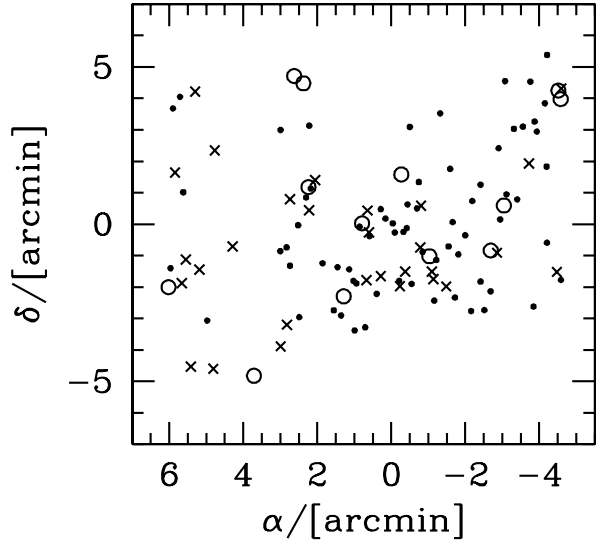}
\caption
{Distribution of member galaxies separated into three clumps according
the one--dimensional KMM test.  The plot is centered on the cluster
center. Dots, open circles, and crosses indicate low--,
intermediate--, and high--velocity clumps, respectively.  }
\label{figkmm}
\end{figure*}

In order to detect possible subclumps in the velocity distribution, we
applied the KMM algorithm (cf. Ashman et al \cite{ash94} and
refs. therein).  By taking the face value of maximum likelihood
statistics, we found a marginal evidence that a mixture of three
Gaussians (of $\mathrm{n_1=13}$, $\mathrm{n_2=70}$, and
$\mathrm{n_3=29}$ members at mean redshift $\mathrm{z_1 = 0.1988}$,
$\mathrm{z_2 = 0.2078}$, and $\mathrm{z_3 = 0.2154}$) is a better
description of velocity distribution (at $91.2\%$ c.l.).  For the
clumps we estimated a velocity dispersion of $\mathrm{\sigma_{v1} =
337}$ \ks, $\mathrm{\sigma_{v2} = 668}$ \ks, and $\mathrm{\sigma_{v3} =
545}$ \ks.  Fig.~\ref{figkmm} shows the spatial distributions of the
three clumps. The second and the third clumps are clearly spatially
segregated.

\subsection{Velocity field}

The above result suggested to investigate the velocity field in more detail.
To this aim, we divided galaxies into a low-- and a high--velocity
sample (with respect to the mean redshift).
Fig.~\ref{figkmm} shows that low-- and high--velocity galaxies are clearly
segregated in the E--W direction, the two distributions being different at the
$99.88\%$ c.l., according to the two--dimensional Kolmogorov--Smirnov
test (hereafter 2DKS--test; cf. Fasano \& Franceschini \cite{fas87},
as implemented by Press et al. \cite{pre92}).

\begin{figure*}
\hspace{1cm}
\vspace{-8cm}
\includegraphics[width=16cm]{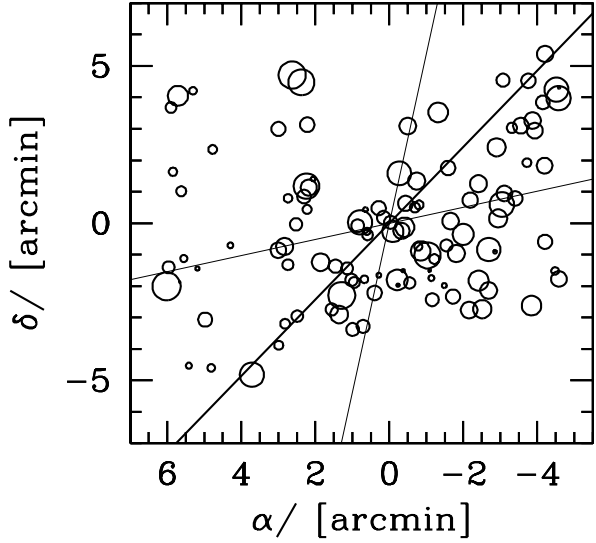}
\caption
{Spatial distribution of member galaxies: the larger the circle, the
smaller is the radial velocity.  The figure is centered on the cluster
center. The solid and the thin lines indicate the position angle of
the cluster velocity gradient and relative errors, respectively.}
\label{figgrad}
\end{figure*}

We then looked for a possible velocity gradient by means of a
multi--linear fit (e.g.,  implemented by NAG Fortran Workstation
Handbook, \cite{nag86}) to the observed velocity field (cf. also den
Hartog \& Katgert \cite{den96}; Girardi et al. \cite{gir96}).  We
found marginal evidence (c.l. $95.2\%$) for the presence of a velocity
gradient in the direction SE--NW, at position
angle PA$=141^{+29}_{-37}$ degrees (cf. Fig.~\ref{figgrad}).
In order to assess the significance of the velocity gradient, we
performed 1000 Monte Carlo simulations by randomly shuffling the
galaxy velocities and determined the coefficient of multiple
determination ($R^2$) for each of them. We then defined the
significance of the velocity gradient as the fraction of times in
which the $R^2$ of the simulation was smaller than the observed $R^2$.

\subsection{3D substructure analysis}

In order to check for the presence of substructure, we combined
velocity and position information by computing the
$\Delta$--statistics \footnote{For each galaxy, the deviation $\delta$
is defined as $\mathrm{\delta^2 = (11/\sigma^2)[(\overline v_l - \overline
v)^2+(\sigma_l - \sigma)^2]}$, where subscript l denotes the average
over the 10 neighbours of the galaxy.  $\Delta$ is the sum of the
$\delta$ of the individual galaxies.} devised by Dressler \& Schectman
(\cite{dre88}) .  We found a value of 162 for the $\Delta$ parameter,
which gives the cumulative deviation of the local kinematical
parameters (velocity mean and velocity dispersion) from the global
cluster parameters.  The significance of substructure was checked by
running 1000 Monte Carlo simulations, randomly shuffling the galaxy
velocities, obtaining a significance level of $98.7\%$.

This indicates that the cluster has a complex structure.

\begin{figure*}
\hspace{1cm}
\vspace{-8cm}
\includegraphics[width=16cm]{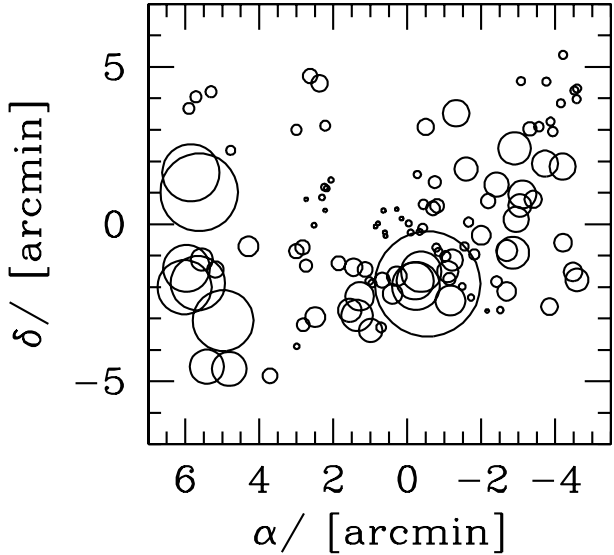}
\caption
{Spatial distribution on the sky of the 112 cluster members, each marked
by a circle: the larger the circle, the larger is the deviation
$\delta$ of the local parameters from the global cluster
parameters. The figure shows evidence for substructure according to
the Dressler \& Schectman test.
The plot is centered on the cluster center. 
}
\label{figds}
\end{figure*}

The technique by Dressler \& Schectman does not allow a direct
identification of galaxies belonging to the detected substructure;
however it can roughly identify the positions of substructures.
To this aim, in Fig.~\ref{figds} the galaxies are marked by circles
whose diameter is proportional to the deviation $\delta$ of the
individual parameters (position and velocity) from the mean cluster
parameters.

A group of galaxies with high velocity is the likely cause of large
values of $\delta$ in the external East cluster region.  The other
possible substructure lies in the well sampled cluster region, closer
than $1$ arcmin SW to the cluster center .

\section{Galaxy populations and the hot gas} 

\subsection{Luminosity and colour segregation}
 
In order to unravel a possible luminosity segregation, we divided the
sample in a low and a high--luminosity subsamples by using the median
V--magnitude$=19.46$ (53 and 54 galaxies, respectively), and applied them
the standard means test and F--test (Press et al. \cite{pre92}).
We found no significant difference between the two samples.

\begin{figure*}
\hspace{1cm}
\vspace{-8cm}
\includegraphics[width=16cm]{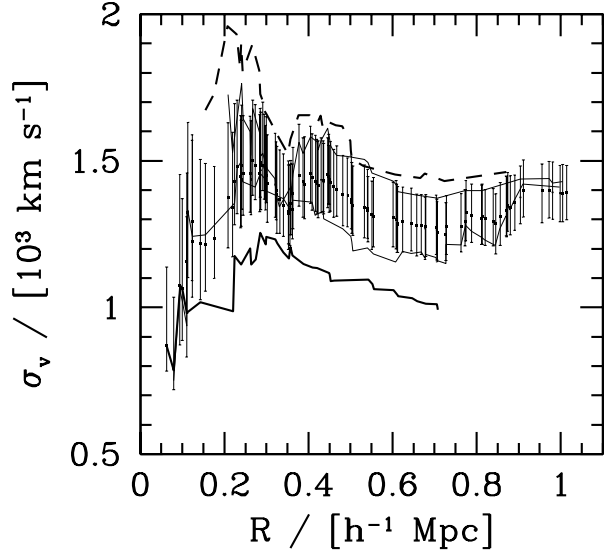}
\caption
{Integrated LOS velocity dispersion profile as in Fig.~\ref{figprof}.
The solid and dashed lines give the profiles of redder and bluer
galaxies, respectively.  The faint lines give the profiles of bright
and faint galaxies (upper and lower lines, respectively).  }
\label{figprofcol}
\end{figure*}

\begin{figure*}
\hspace{1cm}
\vspace{-8cm}
\includegraphics[width=16cm]{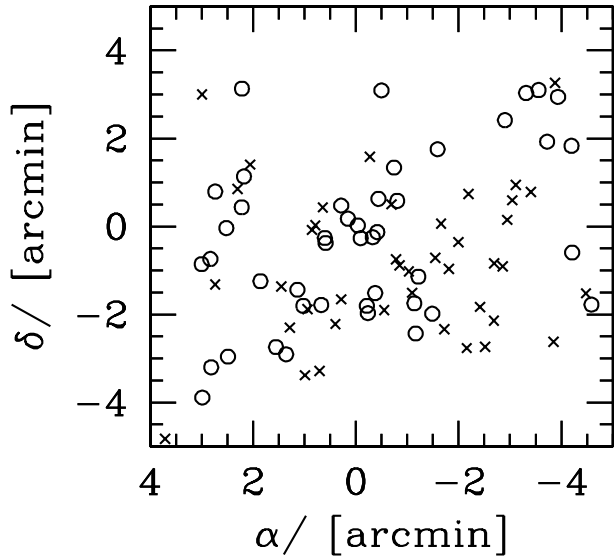}
\caption
{Spatial distribution of the 85 member galaxies having B--R colour
information: circles and crosses give redder and bluer galaxies,
respectively. The plot is centered on the cluster center.
}
\label{figcol}
\end{figure*}

We also looked for possible colour segregation, by dividing the sample
in a blue and a red subsamples relative to the median colour B--R$=2.32$
of the spectroscopic sample (41 and 44 galaxies, respectively).  There
is a slight differences in the peaks of the velocity distributions,
$\mathrm{z_{blue}=0.2076}$ and $\mathrm{z_{red}=0.2095}$, which gives a marginal
probability of difference according to the means test ($93.5\%$).
The velocity dispersions are different being
$\mathrm{\sigma_{v,blue}=1462^{+158}_{-145}}$ \ks larger than
$\mathrm{\sigma_{v,red}=993^{+126}_{-88}}$ \ks (at the $98.6\%$ c.l., according
to the F--test), see also Fig.~\ref{figprofcol}.
Moreover, the two subsamples differ in the distribution of galaxy
positions ($97.4\%$ according to the 2DKS test), 
with blue galaxies lying mainly in the SW--region (cf. Fig.~\ref{figcol}).

\begin{figure*}
\hspace{1cm}
\vspace{-8cm}
\includegraphics[width=16cm]{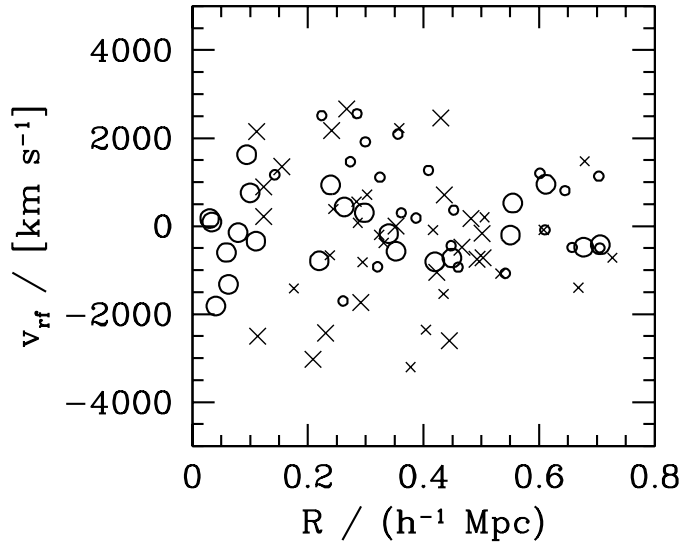}
\caption
{Velocity in the cluster rest--frame vs. (projected) clustercentric distance
for the 85 members having B--R colour information. 
Large circles, small circles, small crosses, and large crosses
indicate galaxies belonging to the four quartiles of the colour distribution,
from reddest to bluest galaxies. 
}
\label{figvdcol}
\end{figure*}

The reddest galaxies (B--R$>2.43$) are characterised by a still
smaller velocity dispersion (732 \kss) and the bluest (B--R$<2.19$)
by a larger velocity dispersion (1689 \kss).  Fig.~\ref{figvdcol}
plots velocities vs. clustercentric distance for the four quartiles of
the colour distribution: the reddest galaxies appear well concentrated
around the mean cluster velocity. From a more quantitative point of
view, there is a significant negative correlation between the B--R
colour and the absolute value of the velocity in the cluster rest
frame $\mathrm{|v_{rf}|}$ at the $99.3\%$ c.l. (for a Kendall correlation
coefficient of $-0.18$).

\begin{figure*}
\centering
\includegraphics[width=0.8\textwidth]{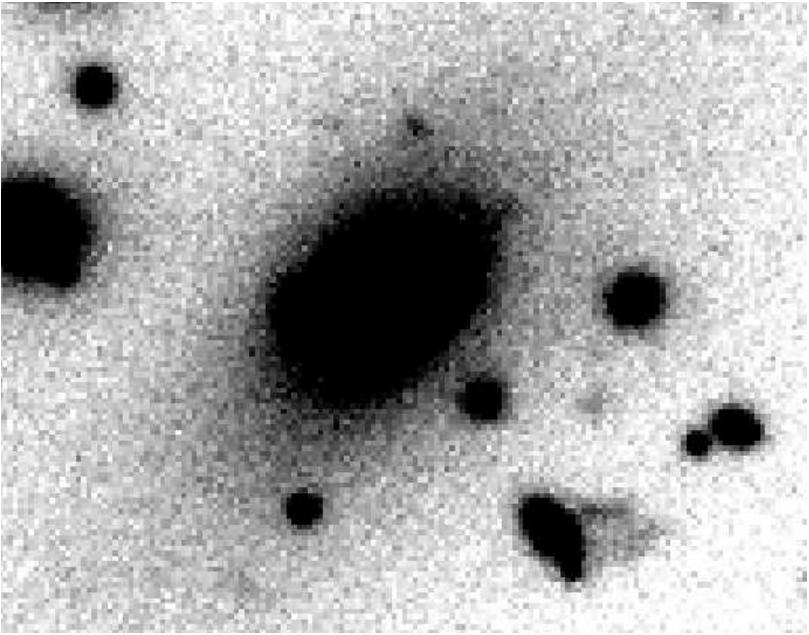}
\caption
{V--band image of cD galaxy ($0.7^\prime \times 0.5^\prime$), with North at top and East to left.}
\label{cD}
\end{figure*}

The velocity of the cD galaxy (z$=0.2095$) shows no evidence of
peculiarity according to the Indicator test by Gebhardt \& Beers
(\cite{geb91}). Moreover, the cD galaxy shows an elongation in the
NE--SW direction (cf. Fig.~\ref{cD}, see La Barbera et al. \cite{lab02}, in
preparation, for more details)

\subsection{Substructures with photometric data}

We performed a two--dimensional analysis to detect subclumps using the
photometric sample of galaxies for which we have B--R colours. This
allows us to take advantage of the larger size of the photometric
sample compared to the spectroscopic one.
Galaxies were selected within $\pm0.5$ mag of the B--R vs. R
colour--magnitude relation and within the completeness limit magnitude
R$=22$ mag.
The colour--magnitude relation was determined on the N=85 spectroscopically 
confirmed cluster members, for which we have B--R colours.
The above selection leads to 3.68 $<$ (B--R) + 0.097 R $<$ 4.68, with
a sample size of N=392 galaxies.

\begin{figure*}
\centering
\includegraphics[width=0.8\textwidth]{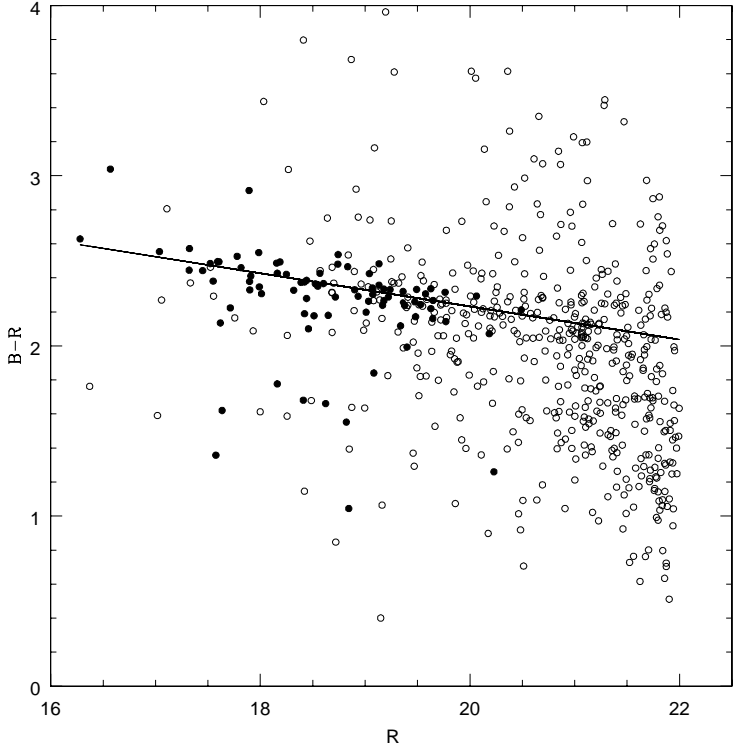}
\caption
{B--R vs. R colour--magnitude diagram for galaxies in the central field of 
ABCG\,209.
Filled circles: spectroscopically confirmed cluster members; open
circles: all galaxies within the completeness limit.}
\label{CM}
\end{figure*}

Fig.~\ref{CM} shows the B--R vs. R colour--magnitude diagram for the
596 galaxies detected in the R--band in the ABCG\,209 central
field. The sample includes also the 85 member galaxies with known
redshift. The colour--magnitude relation was derived by using a linear
regression based on the biweight estimator (see Beers et al. 1990).

\begin{figure*}
\centering
\includegraphics[width=7cm]{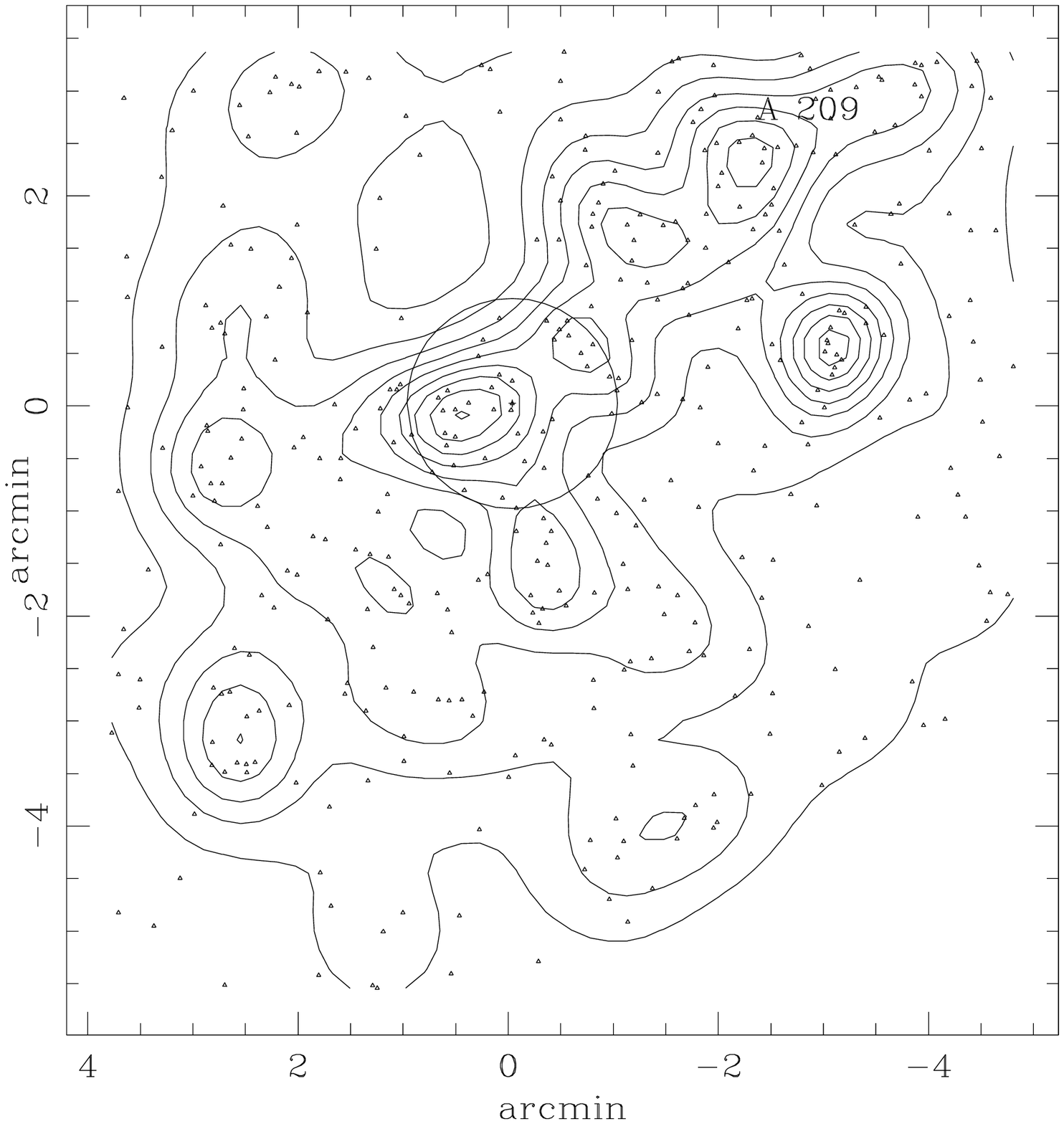}
\includegraphics[width=7cm]{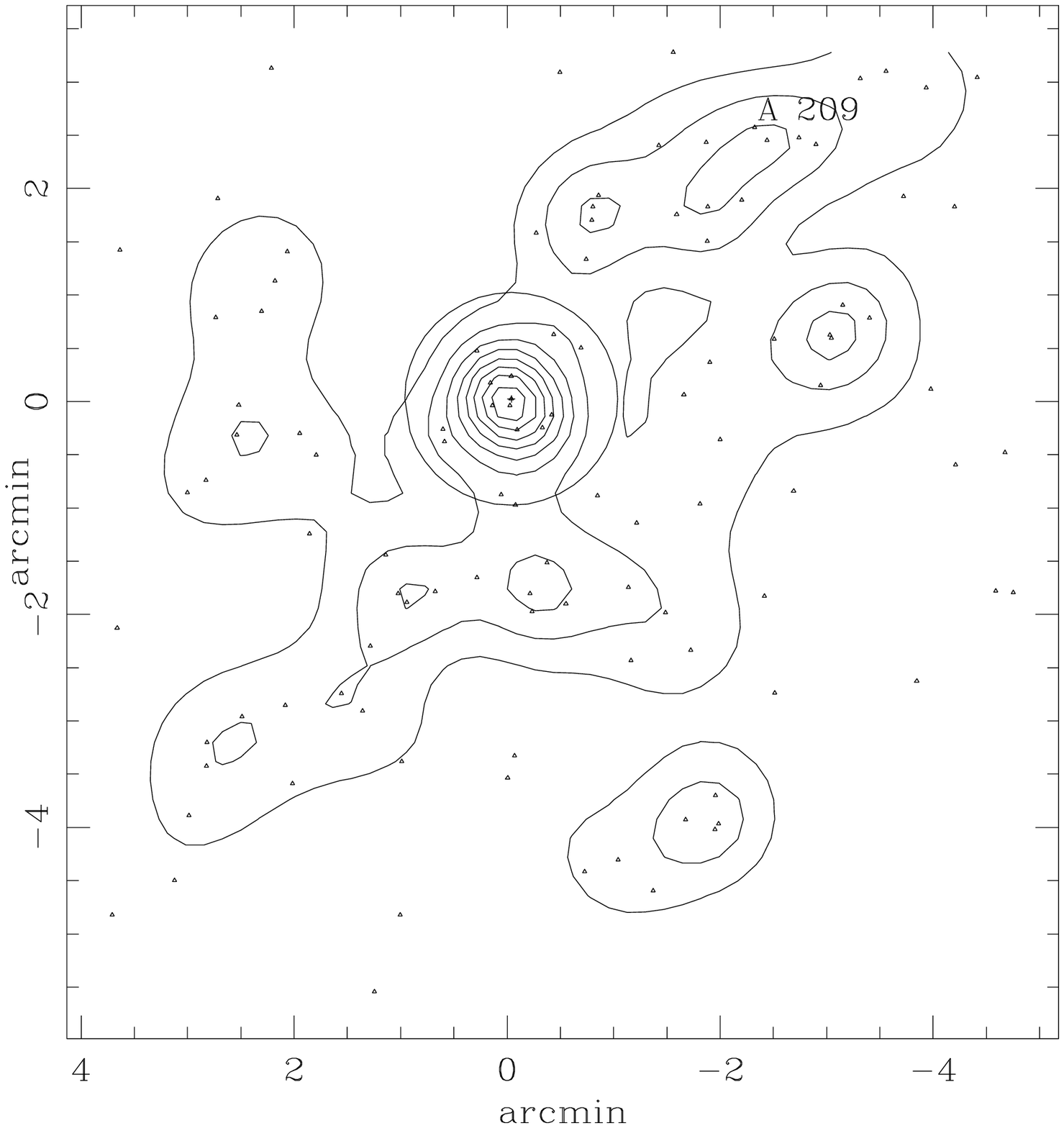}
\includegraphics[width=7cm]{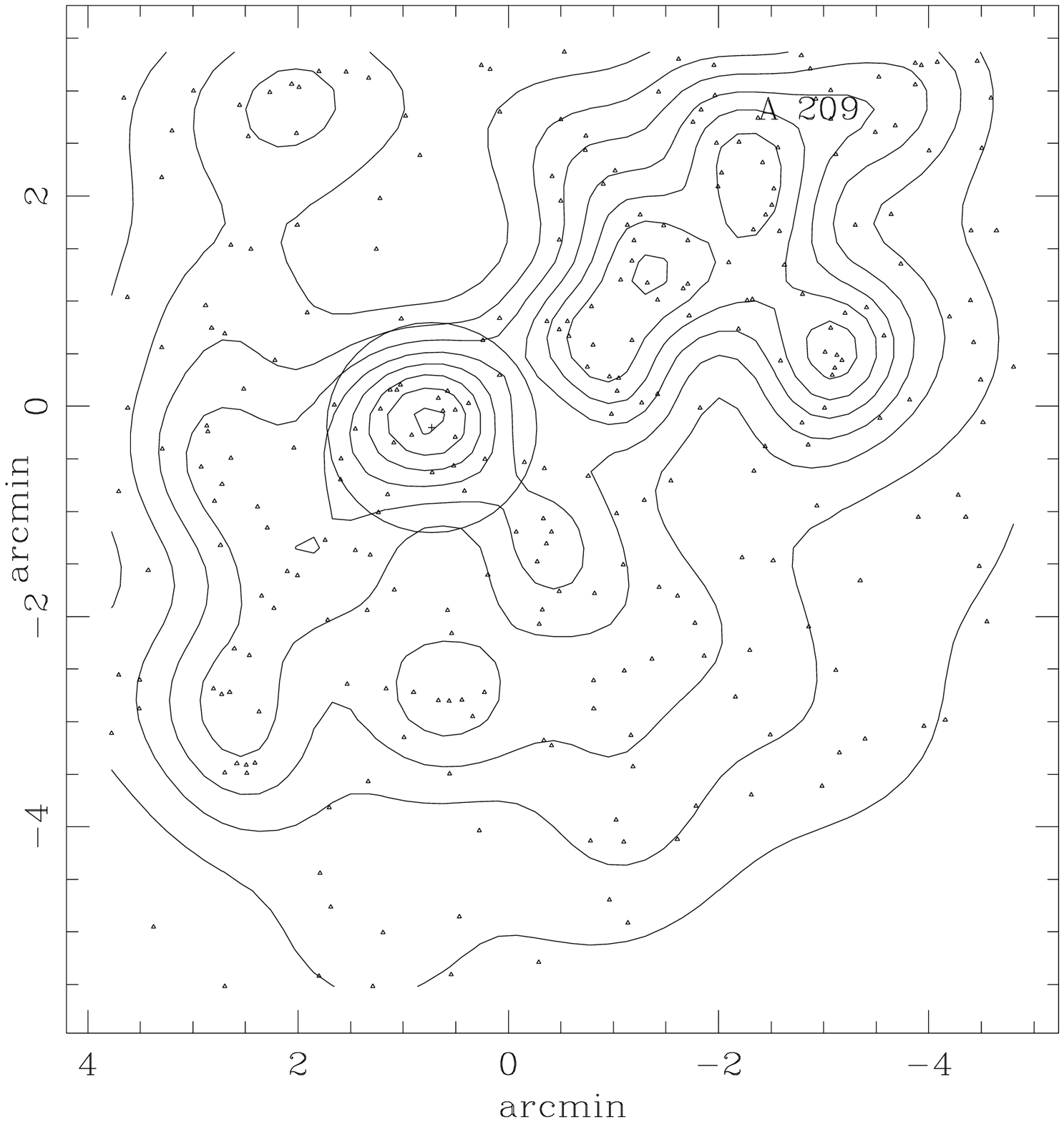}
\caption
{Spatial distribution on the sky and relative isodensity contour map
of likely cluster members selected on the basis of the B--R
colour--magnitude relation, recovered by adaptive--kernel method
(cf. Pisani \cite{pis93},\cite{pis96}).  The plots are centered on the
cluster center.  Top panel: the 392 selected galaxies (see text).  The
1 arcmin circle is centered on the cD galaxy.  Central panel: bright
(R$<19.5$) galaxies. The 1 arcmin circle is again centered on the cD
galaxy.  Bottom panel: fainter (R$>19.5$) galaxies. The 1 arcmin
circle is centered on the secondary X--ray peak (cf. Sect.~5.3). In
the figures the isodensity contours are spaced by 1 $\cdot$ 10$^{-4}$
galaxies/arcsec$^2$ in the top panel and by 5 $\cdot$ 10$^{-5}$
galaxies/arcsec$^2$ in the central and bottom panels.}
\label{fig2d}
\end{figure*}

Fig.~\ref{fig2d} -- upper panel -- shows the projected
galaxy distribution is clustered and elongated in the SE--NW
direction. Notice that there is no clear clump of galaxies perfectly
centered on the cD galaxy, whose position is coincident with the
center determined in Sect.~4.1 from redshift data only.
In order to investigate this question, we noted that galaxies with redshift
data are brighter than 22 R--mag, being the $\sim 90\%$ of them
brighter than 19.5 mag.  We give the distributions of 110 galaxies and
282 galaxies brighter and fainter than 19.5 mag, respectively, in
central and lower panels (Fig.~\ref{fig2d}).  Brighter galaxies are
centered around the cD, while fainter galaxies show some clumps aligned
in the SE--NW direction; in particular the main clump, Eastern with
respect the cD galaxies, coincides with the secondary peak found in
our analysis of Chandra X--ray data (see below).

\subsection{Analysis of X--ray data} 

\begin{figure*}
\centering
\caption
{Optical image of ABCG\,209 with, superimposed, the X--ray contour
levels of the Chandra image, with North at top and East to left. The
ellipses represent the two substructures detected by Wavdetect. The
primary X--ray peak coincides with the position of the cD galaxy and
thus with the peak of the bright galaxy distribution
(cf. Fig.~\ref{fig2d} middle panel).  The secondary X--ray peak
coincides with the main peak of the faint galaxy distribution
(cf. Fig.~\ref{fig2d} lower panel).}

\label{isofote1}
\end{figure*}

The X--ray analysis of ABCG\,209 was performed by using the Chandra
ACIS--I observation 800030 (exposure ID\#522) stored in the Chandra
archive\footnote{http://asc.harvard.edu/cda/}. 
The pointing has an exposure time of 10 ksec.

Data reduction was performed by using the package CIAO\footnote{CIAO is
freely available at http://asc.harvard.edu/ciao/}(Chandra Interactive
Analysis of Observations). First, we removed events from the level 2
event list with a status not equal to zero and with grades one, five
and seven. Then, we selected all events with energy between 0.3 and 10
keV. In addition, we cleaned bad offsets and examined the data on a
chip by chip basis, filtering out bad columns and removing times when
the count rate exceeded three standard deviations from the mean count
rate per 3.3 second interval. We then cleaned each chip for flickering
pixels, i.e. times where a pixel had events in two sequential 3.3
second intervals. The resulting exposure time for the reduced data is
9.86 ks.

The temperature of the ICM was computed extracting the spectrum of the
cluster within a circular aperture of 3 arcminutes radius around the
cluster center. Freezing the absorbing galactic hydrogen column
density at 1.64 10$^{20}$ cm$^{-2}$, computed from the HI maps by
Dickey \& Lockman (\cite{dic90}), we fitted a Raymond--Smith spectrum using the
CIAO package Sherpa with a $\chi^{2}$ statistics.  We found a best
fitting temperature of $\mathrm{T_X=10.2^{+1.4}_{-1.2}}$ keV.

In order to detect possible substructures in ABCG\,209 we ran the task
CIAO/Wavdetect on a subimage of 1600 by 1600 pixels containing the
cluster. The task was run on different scales in order to search for
substructure with different sizes. The significance threshold was set
at $10^{-6}$. 

The results are shown in Fig.~\ref{isofote1}. Two ellipses are plotted
representing two significant substructures found by Wavdetect. 
The principal one, located at $\alpha$= 01 31 52.7 and $\delta$= -13
36 41, is centered on the cD galaxy, the left one is a secondary
structure located at $\alpha$= 01 31 55.7 and $\delta$= -13 36 54,
about 50 arcseconds ($\sim$120 $\mathrm{h^{-1}}$ kpc) East of the cD (cf. also
Fig.~\ref{fig2d}).
The secondary clump is well coincident with the Eastern clump detected
by Rizza et al.  (\cite{riz98}) by using ROSAT HRI X--ray data, while
we do not find any significant substructure corresponding to their
Western excess.

\section{Formation and evolution of ABCG\,209}

The value we obtained for the LOS velocity dispersion
$\mathrm{\sigma_v\sim 1400}$ \ks (Sect.~4.1) is high compared to the
typical, although similar values are sometimes found in clusters at
intermediate redshifts (cf. Fadda et al. \cite{fad96}; Mazure et
al. \cite{maz96}; Girardi \& Mezzetti \cite{gir01}).  In Sect.~4.2 we
show that this high value of $\mathrm{\sigma_v}$ is not due to obvious
interlopers: also a very restrictive application of the ``shifting
gapper'' or the rejection of minor peaks in galaxy density lead to
only a slightly smaller value of $\mathrm{\sigma_v\sim1250}$ \kss.
The high global value of $\mathrm{\sigma_v=1250}$--$1400$ \ks is
consistent with the high value of $\mathrm{T_X \sim 10}$ keV coming
from the X--ray analysis (which, assuming $\mathrm{\beta_{spec}=1}$,
would correspond to $\mathrm{\sigma_v\sim1300}$ \ks,
cf. Fig.~\ref{figprof}) and with the high value of
$\mathrm{L_X(0.1-2.4\ keV)\sim14\;h_{50}^{-2}\;10^{44}}$ erg
$\mathrm{s^{-1}}$ (Ebeling et al. \cite{ebe96}; cf. with
$\mathrm{L_X}$--$\mathrm{\sigma_v}$ relation by e.g., Wu et
al. \cite{wu99}; Girardi \& Mezzetti \cite{gir01}).

Therefore, on the basis of the global properties only, one could
assume that ABCG\,209 is not far from dynamical equilibrium and rely on
large virial mass estimate here computed
$\mathrm{M(<1.78}$ \hh$ )=2.25^{+0.63}_{-0.65}\times10^{15}$ \msun or
$\mathrm{M(<1.59}$ \hh$ )=1.62^{+0.48}_{-0.46}\times10^{15}$ \msun.

On the other hand, the analysis of the integral velocity dispersion
profile shows that a high value of $\mathrm{\sigma_v}$ is already reached
within the central cluster region of 0.2--0.3 \hh.
This suggests the possibility that a mix of clumps at different mean
velocity causes the high value of the velocity dispersion.

A deeper analysis shows that ABCG\,209 is currently undergoing a
dynamical evolution.
We find evidence for a preferential SE--NW direction as indicated
by: a) the presence of a velocity gradient; b) the elongation in the
spatial distribution of the colour--selected likely cluster members; c)
the elongation of the X--ray contour levels in the Chandra image; d) the
elongation of the cD galaxy.

In particular, velocity gradients are rarely found in clusters (e.g., 
den Hartog \& Katgert \cite{den96}) and could be produced by rotation,
by presence of internal substructures,
and by presence of other structures on larger
scales such as nearby clusters, surrounding superclusters, or filaments
(e.g., West \cite{wes94}; Praton \& Schneider \cite{pra94}). 

The elongation of the cD galaxies, aligned along
the major axis of the cluster and of the surrounding LSS (e.g.,
Binggeli \cite{bin82}; Dantas et al. \cite{dan97}; Durret et
al. \cite{dur98}), can be explained if BCMs form by
coalescence of the central brightest galaxies of the merging
subclusters (Johnstone et al. \cite{joh91}).

Other evidence that this cluster is far from dynamical equilibrium
comes out from deviation of velocity distribution from Gaussian, 
spatial and kinematical segregation of members with different B--R colour, and
evidence of substructure as given by the Dressler--Schectman test
(\cite{dre88}).  In particular, although a difference in $\mathrm{\sigma_v}$
between blue and red 
members is common in all clusters (e.g., Carlberg
et al. \cite{car97}), a displacement in mean velocity or in position
center is more probably associated with a situation of non equilibrium
(e.g., Bruzendorf \& Meusinger \cite{bru99}).

Possible subclumps, if any, cannot be
easily separated by using 112 cluster members only. The separation in
three clumps by using one--dimensional KMM test and the visual
inspection of Dressler--Schectman map of Fig.~\ref{figds} represent
only two tentative approaches (cf. Sects.~4.3 and 4.5).

A further step in the detection of possible subclumps is achieved by
using the two--dimensional complete sample of all $\sim400$
colour--selected cluster members. In fact, several clumps appear well
aligned in the SE--NW direction, and, in particular, the main clump 
revealed by faint galaxies (R$>19.5$ mag) coincides with
the secondary X--ray peak as found in our analysis of Chandra data
(cf. also Rizza et al. \cite{riz98}). This result suggests that one or
several minor clumps are merging along the SE--NW direction, with a
main clump hosting the cD galaxy.  In particular, the presence of a
significant velocity gradient suggests that the merging plane is not
parallel to the plane of the sky.

The presence of a secondary clump as a well distinct unit in
X--ray data suggests that this clump and maybe the whole cluster might be
in a pre--merger phase.

The strong luminosity segregation found for colour--selected galaxies
could however suggests a possible alternative dynamical scenario.  In
fact, very appealingly, galaxies of different luminosity could trace
the dynamics of cluster mergers in a different way.  A first evidence
was given by Biviano et al.  (\cite{biv96}): they found that the two
central dominant galaxies of the Coma cluster are surrounded by
luminous galaxies, accompanied by the two main X--ray peaks, while the
distribution of faint galaxies tend to form a structure not centered
with one of the two dominant galaxies, but rather coincident with a
secondary peak detected in X--ray.  The observational scenario of
ABCG\,209 has some similarities with the situation of Coma: bright
galaxies (R$<19.5$ mag) are concentrated around the cD galaxy, which
coincides with the primary X--ray peak, while faint galaxies
(R$>19.5$ mag) show several peaks, the main of which, Eastern with
respect the cD galaxy, is confirmed by the position of the secondary
X--ray peak.
Therefore, following Biviano et al., we might speculate that the
merging is in an advanced phase, where faint galaxies trace the
forming structure of the cluster, while more luminous galaxies still
trace the remnant of the core--halo structure of a pre--merging clump,
which could be so dense to survive for a long time after the merging
(as suggested by numerical simulations Gonz\'alez--Casado et
al. \cite{gon94}).  
An extended Radio--emission would support an advanced merging phase,
but its presence is still uncertain due to the existence of strong
discrete sources (Giovannini et al. \cite{gio99}). The comparison of
present Radio--image with our galaxy distribution shows that the
diffuse source is located around the central cD galaxy with an
extension toward North.

Unfortunately, redshift data are available only for luminous galaxies
and we cannot investigate the nature of subclumps inferred from the
two--dimensional distribution of faint galaxies.
At the same time, available X--ray data are not deep enough to look
for possible variations of temperature in the region of interest.

\section{Summary}

In order to study the internal dynamics of the rich galaxy cluster
ABCG\,209, we obtained spectra for 159 objects in the cluster region
based on MOS observations carried out at the ESO New Technology
Telescope.  Out of these spectra, we analysed 119 galaxies: 
112 turn out to be cluster members, 1 is foreground and 6 are
background galaxies.

ABCG\,209 appears as a well isolated peak in the redshift distribution
centered at $z=0.209$, characterized by a very high value of the LOS
velocity dispersion: $\mathrm{\sigma_v=1250}$--$1400$ \kss, that
results in a virial mass of $\mathrm{M=1.6}$--$2.2\times 10^{15}$\msun
within R$\mathrm{_{vir}}$.  The analysis of the velocity dispersion
profile show that such high value of $\mathrm{\sigma_v}$ is already
reached in the central cluster region ($<0.2$--0.3 \hh).

The main results of the present study may be summarised as follows.

\begin{itemize}

\item ABCG\,209 is characterised by a preferential SE--NW direction as
indicated by: a) the presence of a velocity gradient in the velocity
field; b) the elongation in the spatial distribution of colour--selected
likely cluster members; c) the elongation of the X--ray contour levels in 
the Chandra image; d) the elongation of the cD galaxy.

\item We find significant deviations of velocity distribution from Gaussian.

\item Red and blue members are spatially and kinematically segregated.

\item There is significant evidence of substructure, as shown by the
Dressler \& Schectman test.

\item The two--dimensional distribution of the colour--selected likely
members shows a strong luminosity segregation: bright galaxies
$\mathrm{R<19.5}$ are centered around the cD galaxy, while faint galaxies
$\mathrm{R>19.5}$ show some clumps. The main one, Eastern with respect to the
cD galaxy, is well coincident with the secondary X--ray peak.

\end{itemize}

This observational scenario suggests that ABCG\,209 is presently
undergoing strong dynamical evolution. Present results suggest the
merging of two or more subclumps along the SE--NW direction in a plane
which is not parallel to the plane of sky, but cannot discriminate
between two alternative pictures.  The merging might be in a very
early dynamical status, where clumps are still in the pre--merging
phase, or in a more advanced status, where luminous galaxies trace the
remnant of the core--halo structure of a pre--merging clump hosting
the cD galaxy.

The connection between the dynamics and the properties of galaxy populations
will be discussed in a forthcoming paper.

\begin{acknowledgements}
We thank Rafael Barrena, Andrea Biviano, Francesco La Barbera, Michele
Massarotti and Massimo Ramella, for useful discussions and
suggestions. We thank the anonymous referee for useful remarks and
comments.  A. M. thanks Massimo Capaccioli for the hospitality at the
Osservatorio Astronomico di Capodimonte and Francesca Matteucci for
support during this work. This research has made use of the NASA/IPAC
extragalactic Database (NED), which is operated by the Jet Propulsion
Laboratory, California Institute of Technology, under contract to the
National Aeronautics and Space Administration.  the X--ray data used
in this work have been obtained from the Chandra data archive at the
NASA Chandra X--ray center. This work has been partially supported by
the Italian Ministry of Education, University, and Research (MIUR)
grant COFIN2001028932: {\it Clusters and groups of galaxies, the
interplay of dark and baryonic matter}, and by the Italian Space
Agency (ASI).
\end{acknowledgements}

\begin{table}
        \caption[]{Spectroscopic data.}
         \label{catalogue}
              $$ 
           \begin{array}{c c c c c c c}
            \hline
            \noalign{\smallskip}
            \hline
            \noalign{\smallskip}

\mathrm{ID} & {\mathrm{\alpha}} & \mathrm{\delta}  & \mathrm{V} & B-R  & z & 
\mathrm{\Delta} z\\
            \hline
            \noalign{\smallskip}   
  1 & 01\ 31\ 33.81 & -13\ 32\ 22.9 & 18.32 & .... & 0.2191 & 0.0003 \\
  2 & 01\ 31\ 33.82 & -13\ 38\ 28.5 & 19.27 & 2.43 & 0.2075 & 0.0002 \\
  3 & 01\ 31\ 33.86 & -13\ 32\ 43.4 & 19.46 & .... & 0.1998 & 0.0003 \\
  4 & 01\ 31\ 34.11 & -13\ 32\ 26.8 & 19.90 & .... & 0.2000 & 0.0004 \\
  5 & 01\ 31\ 34.26 & -13\ 38\ 13.1 & 20.17 & 2.24 & 0.2145 & 0.0005 \\
  6 & 01\ 31\ 35.37 & -13\ 31\ 18.9 & 17.36 & .... & 0.2068 & 0.0004 \\
  7 & 01\ 31\ 35.37 & -13\ 37\ 17.4 & 18.58 & 2.38 & 0.2087 & 0.0002 \\
  8 & 01\ 31\ 35.42 & -13\ 34\ 52.0 & 19.10 & 2.37 & 0.2073 & 0.0002 \\
  9 & 01\ 31\ 35.61 & -13\ 32\ 51.3 & 20.17 & .... & 0.2090 & 0.0002 \\
 10 & 01\ 31\ 36.51 & -13\ 33\ 45.2 & 19.83 & 2.36 & 0.2072 & 0.0004 \\
 11^{\mathrm{a}} & 01\ 31\ 36.54 & -13\ 37\ 56.1 & 19.89 & 1.37 & 0.2620 & 0.0005 \\
 12 & 01\ 31\ 36.76 & -13\ 33\ 26.1 & 21.10 & 2.21 & 0.2064 & 0.0004 \\ 
 13 & 01\ 31\ 36.87 & -13\ 39\ 19.3 & 19.73 & 2.26 & 0.2039 & 0.0002 \\
 14^{\mathrm{a}} & 01\ 31\ 37.19 & -13\ 30\ 04.3 & 20.65 & .... & 0.3650 & 0.0003 \\
 15 & 01\ 31\ 37.23 & -13\ 32\ 09.9 & 20.03 & .... & 0.2084 & 0.0003 \\
 16 & 01\ 31\ 37.38 & -13\ 34\ 46.3 & 18.71 & 2.35 & 0.2134 & 0.0003 \\
 17^{\mathrm{a}} & 01\ 31\ 37.99 & -13\ 36\ 01.6 & 21.38 & 2.67 & 0.3997 & 0.0004 \\ 
 18 & 01\ 31\ 38.06 & -13\ 33\ 35.9 & 19.87 & 2.49 & 0.2073 & 0.0002 \\
 19 & 01\ 31\ 39.06 & -13\ 33\ 39.9 & 19.64 & 2.33 & 0.2120 & 0.0002 \\
 20 & 01\ 31\ 38.69 & -13\ 35\ 54.7 & 20.11 & 2.17 & 0.2084 & 0.0003 \\
 21 & 01\ 31\ 39.89 & -13\ 35\ 45.1 & 18.85 & 1.68 & 0.2073 & 0.0002 \\
 22 & 01\ 31\ 40.05 & -13\ 32\ 08.9 & 19.51 & .... & 0.2096 & 0.0003 \\
 23 & 01\ 31\ 40.17 & -13\ 36\ 06.2 & 19.18 & 2.18 & 0.1995 & 0.0002 \\
 24 & 01\ 31\ 40.58 & -13\ 36\ 32.9 & 19.30 & 2.18 & 0.2052 & 0.0003 \\
 25 & 01\ 31\ 40.76 & -13\ 34\ 17.0 & 18.58 & 2.33 & 0.2051 & 0.0002 \\   
 26 & 01\ 31\ 40.94 & -13\ 37\ 36.5 & 19.10 & 1.66 & 0.2181 & 0.0008 \\
 27 & 01\ 31\ 41.63 & -13\ 37\ 32.3 & 19.39 & 2.28 & 0.2004 & 0.0002 \\
 28 & 01\ 31\ 41.64 & -13\ 38\ 50.2 & 19.22 & 1.55 & 0.2063 & 0.0003 \\
 29 & 01\ 31\ 42.35 & -13\ 39\ 26.0 & 19.07 & 2.19 & 0.2051 & 0.0004 \\
 30 & 01\ 31\ 42.76 & -13\ 38\ 31.6 & 19.76 & 2.31 & 0.2034 & 0.0002 \\
 31 & 01\ 31\ 42.77 & -13\ 35\ 26.6 & ..... & .... & 0.2062 & 0.0007 \\
 32 & 01\ 31\ 43.69 & -13\ 35\ 57.8 & 20.73 & 2.30 & 0.2076 & 0.0003 \\
 33 & 01\ 31\ 43.81 & -13\ 39\ 27.7 & 20.40 & 2.14 & 0.2064 & 0.0002 \\
 34 & 01\ 31\ 44.47 & -13\ 37\ 03.2 & 19.96 & 1.99 & 0.2027 & 0.0003 \\
 35 & 01\ 31\ 45.24 & -13\ 37\ 39.7 & 18.39 & 2.23 & 0.2060 & 0.0002 \\
 36 & 01\ 31\ 45.61 & -13\ 39\ 02.0 & 19.88 & 2.29 & 0.2087 & 0.0004 \\
 37 & 01\ 31\ 45.87 & -13\ 36\ 38.0 & 18.72 & 2.31 & 0.2066 & 0.0002 \\
 38 & 01\ 31\ 46.15 & -13\ 34\ 56.6 & 18.56 & 2.46 & 0.2084 & 0.0002 \\
 39 & 01\ 31\ 46.33 & -13\ 37\ 24.5 & 20.29 & 2.22 & 0.2105 & 0.0003 \\
 40 & 01\ 31\ 46.58 & -13\ 38\ 40.9 & 19.02 & 2.32 & 0.2167 & 0.0002 \\
 41 & 01\ 31\ 47.26 & -13\ 33\ 10.3 & 19.72 & .... & 0.2038 & 0.0004 \\
 42 & 01\ 31\ 47.69 & -13\ 37\ 50.4 & 18.18 & 2.44 & 0.2125 & 0.0003 \\
 43 & 01\ 31\ 47.91 & -13\ 39\ 07.9 & 18.66 & 2.41 & 0.2097 & 0.0003 \\
              \noalign{\smallskip}			    
            \hline					    
            \noalign{\smallskip}			    
            \hline					    
         \end{array}
     $$ 
         \end{table}
\addtocounter{table}{-1}
\begin{table}
          \caption[ ]{Continued.}
     $$ 
           \begin{array}{c c c c c c c}
            \hline
            \noalign{\smallskip}
            \hline
            \noalign{\smallskip}

\mathrm{ID} & {\mathrm{\alpha}} & \mathrm{\delta}  & \mathrm{V} & B-R  & z & 
\mathrm{\Delta} z\\
            \hline
            \noalign{\smallskip}
 44 & 01\ 31\ 48.01 & -13\ 38\ 26.7 & 19.32 & 2.37 & 0.2161 & 0.0002 \\
 45 & 01\ 31\ 48.20 & -13\ 38\ 12.3 & 20.82 & 2.08 & 0.2188 & 0.0001 \\
 46 & 01\ 31\ 48.45 & -13\ 37\ 43.2 & 20.34 & 2.16 & 0.1979 & 0.0005 \\
 47 & 01\ 31\ 49.21 & -13\ 37\ 35.0 & 19.66 & 2.20 & 0.2039 & 0.0003 \\
 48 & 01\ 31\ 49.38 & -13\ 36\ 06.9 & 20.53 & 2.32 & 0.2133 & 0.0004 \\
 49 & 01\ 31\ 49.47 & -13\ 37\ 26.5 & 18.02 & 1.62 & 0.2140 & 0.0002 \\
 50 & 01\ 31\ 49.64 & -13\ 35\ 21.6 & 19.74 & 2.43 & 0.2061 & 0.0002 \\
 51 & 01\ 31\ 49.84 & -13\ 36\ 11.7 & 20.15 & 2.12 & 0.2123 & 0.0003 \\
 52 & 01\ 31\ 50.43 & -13\ 38\ 35.9 & 19.89 & 2.26 & 0.2111 & 0.0003 \\
 53 & 01\ 31\ 50.64 & -13\ 33\ 36.4 & 18.13 & 2.45 & 0.2064 & 0.0004 \\
 54 & 01\ 31\ 50.89 & -13\ 36\ 04.2 & 18.77 & 2.54 & 0.2078 & 0.0002 \\
 55 & 01\ 31\ 50.98 & -13\ 36\ 49.5 & 19.64 & 2.54 & 0.2042 & 0.0003 \\
 56 & 01\ 31\ 51.15 & -13\ 38\ 12.8 & 18.27 & 2.38 & 0.2183 & 0.0002 \\
 57 & 01\ 31\ 51.34 & -13\ 36\ 56.6 & 18.10 & 2.58 & 0.2068 & 0.0002 \\
 58 & 01\ 31\ 51.58 & -13\ 35\ 07.0 & 18.24 & 2.13 & 0.2001 & 0.0002 \\
 59 & 01\ 31\ 51.73 & -13\ 38\ 40.2 & 19.26 & 2.37 & 0.2184 & 0.0002 \\
 60 & 01\ 31\ 51.82 & -13\ 38\ 30.2 & 20.23 & 2.33 & 0.2028 & 0.0005 \\
 61 & 01\ 31\ 52.31 & -13\ 36\ 57.9 & 17.37 & 3.04 & 0.2024 & 0.0002 \\
 62 & 01\ 31\ 52.54 & -13\ 36\ 40.4 & 17.00 & 2.63 & 0.2097 & 0.0002 \\
 63 & 01\ 31\ 53.34 & -13\ 36\ 31.3 & 18.52 & 2.92 & 0.2094 & 0.0002 \\
 64 & 01\ 31\ 53.86 & -13\ 38\ 21.2 & 19.66 & 1.84 & 0.2170 & 0.0003 \\
 65 & 01\ 31\ 53.87 & -13\ 36\ 13.4 & 18.90 & 2.48 & 0.2085 & 0.0002 \\
 66^{\mathrm{a}} & 01\ 31\ 54.09 & -13\ 39\ 39.0 & 21.03 & 2.73 & 0.4538 & 0.0001 \\
 67 & 01\ 31\ 54.33 & -13\ 38\ 54.9 & 20.04 & 2.26 & 0.2083 & 0.0004 \\
 68 & 01\ 31\ 55.12 & -13\ 37\ 04.4 & 18.93 & 2.49 & 0.2118 & 0.0003 \\
 69 & 01\ 31\ 55.18 & -13\ 36\ 57.6 & 18.94 & 2.43 & 0.2150 & 0.0002 \\
 70 & 01\ 31\ 55.34 & -13\ 36\ 15.9 & 19.14 & 1.05 & 0.2169 & 0.0003 \\
 71 & 01\ 31\ 55.47 & -13\ 38\ 28.9 & 19.18 & 2.39 & 0.2144 & 0.0002 \\
 72 & 01\ 31\ 55.61 & -13\ 39\ 58.9 & 20.65 & 1.26 & 0.2097 & 0.0002 \\
 73 & 01\ 31\ 55.95 & -13\ 36\ 40.4 & 18.00 & 1.35 & 0.1999 & 0.0005 \\
 74 & 01\ 31\ 56.22 & -13\ 36\ 46.7 & 18.43 & 1.78 & 0.2098 & 0.0003 \\
 75 & 01\ 31\ 56.58 & -13\ 38\ 34.9 & 20.03 & 2.23 & 0.2117 & 0.0002 \\
 76 & 01\ 31\ 56.78 & -13\ 40\ 04.8 & 20.16 & 2.26 & 0.2098 & 0.0004 \\
 77 & 01\ 31\ 56.91 & -13\ 38\ 30.2 & 18.26 & 2.49 & 0.2102 & 0.0002 \\
 78 & 01\ 31\ 57.38 & -13\ 38\ 08.2 & 19.58 & 2.47 & 0.2107 & 0.0003 \\
 79 & 01\ 31\ 57.99 & -13\ 38\ 59.7 & 19.14 & 2.28 & 0.1973 & 0.0003 \\
 80 & 01\ 31\ 58.28 & -13\ 39\ 36.2 & 19.14 & 2.38 & 0.2056 & 0.0002 \\
 81 & 01\ 31\ 58.68 & -13\ 38\ 04.0 & 20.32 & 2.27 & 0.2093 & 0.0004 \\
 82 & 01\ 31\ 59.10 & -13\ 39\ 26.4 & 18.97 & 2.42 & 0.2104 & 0.0002 \\
 83 & 01\ 32\ 00.34 & -13\ 37\ 56.5 & 20.02 & 2.32 & 0.2056 & 0.0002 \\
 84 & 01\ 32\ 01.18 & -13\ 35\ 17.5 & 19.65 & 2.29 & 0.2172 & 0.0003 \\   
 85^{\mathrm{a}} & 01\ 32\ 01.56 & -13\ 32\ 21.1 & 19.95 & .... & 0.2617 & 0.0003 \\ 
 86 & 01\ 32\ 01.66 & -13\ 35\ 33.8 & 18.36 & 2.49 & 0.2069 & 0.0002 \\

              \noalign{\smallskip}			    
            \hline					    
            \noalign{\smallskip}			    
            \hline					    
         \end{array}
     $$ 
         \end{table}
\addtocounter{table}{-1}
\begin{table}
          \caption[ ]{Continued.}
     $$ 
           \begin{array}{c c c c c c c}
            \hline
            \noalign{\smallskip}
            \hline
            \noalign{\smallskip}

\mathrm{ID} & {\mathrm{\alpha}} & \mathrm{\delta}  & \mathrm{V} & B-R  & z & 
\mathrm{\Delta} z\\
            \hline
            \noalign{\smallskip}
 87 & 01\ 32\ 01.82 & -13\ 33\ 34.0 & 18.60 & 2.52 & 0.2083 & 0.0003 \\ 
 88 & 01\ 32\ 01.84 & -13\ 36\ 15.7 & ....  & 2.33 & 0.2131 & 0.0004 \\
 89 & 01\ 32\ 01.91 & -13\ 35\ 31.2 & 18.65 & .... & 0.1984 & 0.0002 \\
 90 & 01\ 32\ 02.18 & -13\ 35\ 51.0 & 19.06 & 2.10 & 0.2091 & 0.0003 \\
 91 & 01\ 32\ 02.47 & -13\ 32\ 13.0 & 19.90 & .... & 0.1982 & 0.0004 \\
 92 & 01\ 32\ 02.94 & -13\ 39\ 39.4 & 19.48 & 2.48 & 0.2110 & 0.0002 \\
 93 & 01\ 32\ 03.07 & -13\ 36\ 44.0 & 19.95 & 2.33 & 0.2102 & 0.0005 \\
 94 & 01\ 32\ 03.96 & -13\ 35\ 54.4 & 19.92 & 2.33 & 0.2137 & 0.0003 \\
 95 & 01\ 32\ 03.97 & -13\ 38\ 01.2 & 20.31 & 2.16 & 0.2116 & 0.0003 \\
 96 & 01\ 32\ 03.50 & -13\ 31\ 59.1 & 19.65 & .... & 0.1971 & 0.0003 \\
 97 & 01\ 32\ 04.29 & -13\ 39\ 53.9 & 17.82 & 2.55 & 0.2125 & 0.0002 \\
 98 & 01\ 32\ 04.35 & -13\ 37\ 26.3 & 18.36 & 2.50 & 0.2061 & 0.0002 \\
 99 & 01\ 32\ 05.00 & -13\ 40\ 35.1 & 19.77 & 2.34 & 0.2132 & 0.0002 \\
100 & 01\ 32\ 05.02 & -13\ 33\ 42.0 & 20.34 & 2.30 & 0.2087 & 0.0003 \\ 
101 & 01\ 32\ 05.05 & -13\ 37\ 33.3 & 19.28 & 2.35 & 0.2074 & 0.0002 \\
102^{\mathrm{b}} & 01\ 32\ 07.29 & -13\ 37\ 30.3 & 19.06 & 1.63 & 0.1744 & 0.0006 \\
103 & 01\ 32\ 07.98 & -13\ 41\ 31.4 & 19.86 & 2.23 & 0.1997 & 0.0002 \\
104 & 01\ 32\ 10.37 & -13\ 37\ 24.3 & 19.54 & .... & 0.2161 & 0.0002 \\
105 & 01\ 32\ 12.34 & -13\ 34\ 21.1 & 19.74 & .... & 0.2134 & 0.0004 \\
106 & 01\ 32\ 12.51 & -13\ 41\ 18.1 & ..... & .... & 0.2145 & 0.0003 \\
107 & 01\ 32\ 13.19 & -13\ 39\ 46.0 & ..... & .... & 0.2087 & 0.0003 \\
108^{\mathrm{a}} & 01\ 32\ 13.22 & -13\ 31\ 09.1 & 22.34 & .... & 0.2597 & 0.0005 \\
109 & 01\ 32\ 14.04 & -13\ 38\ 08.5 & 17.90 & .... & 0.2175 & 0.0002 \\
110 & 01\ 32\ 14.52 & -13\ 32\ 29.1 & 19.39 & .... & 0.2145 & 0.0002 \\
111 & 01\ 32\ 15.00 & -13\ 41\ 13.9 & ..... & .... & 0.2158 & 0.0003 \\
112 & 01\ 32\ 15.56 & -13\ 37\ 49.2 & 18.94 & .... & 0.2149 & 0.0002 \\
113 & 01\ 32\ 15.84 & -13\ 35\ 41.0 & 20.71 & .... & 0.2124 & 0.0004 \\
114 & 01\ 32\ 16.00 & -13\ 38\ 34.6 & 19.57 & .... & 0.2203 & 0.0003 \\
115 & 01\ 32\ 16.20 & -13\ 32\ 38.8 & 18.97 & .... & 0.2034 & 0.0003 \\
116 & 01\ 32\ 16.98 & -13\ 33\ 01.1 & 20.64 & .... & 0.2117 & 0.0004 \\
117 & 01\ 32\ 16.76 & -13\ 35\ 03.7 & 20.81 & .... & 0.2138 & 0.0004 \\
118 & 01\ 32\ 17.26 & -13\ 38\ 06.1 & 19.74 & .... & 0.2105 & 0.0003 \\
119 & 01\ 32\ 17.48 & -13\ 38\ 42.3 & 19.84 & .... & 0.1963 & 0.0006 \\
            \noalign{\smallskip}			    
            \hline					    
            \noalign{\smallskip}			    
            \hline					    
         \end{array}
     $$ 
\begin{list}{}{}  
\item[$^{\mathrm{a}}$] Background galaxy.
\item[$^{\mathrm{b}}$] Foreground galaxy.
\end{list}
   \end{table}

\end{document}